\documentclass[aps,11pt,onecolumn,preprintnumbers,amsmath,amssymb]{revtex4}

\usepackage[english]{babel}
\usepackage{graphicx}
\usepackage{color}

\begin{document}

\title{Quantum correlations of confined exciton-polaritons}
\author{Aymeric Delteil$^{1*}$, Thomas Fink$^{1*}$, Anne Schade$^{2}$, Sven H\"ofling$^{2,3}$, Christian Schneider$^{2}$ and Ata\c{c} \.Imamo\u{g}lu$^{1}$}
\affiliation{$^1$ Institute of Quantum Electronics, ETH Zurich, 8093 Zurich, Switzerland. \\
$^2$ Technische Physik, Universit\"at W\"urzburg, W\"urzburg, Germany.  \\
$^3$ SUPA, School of Physics and Astronomy, University of St Andrews, St Andrews, UK. \\
$^*$ These authors contributed equally to this work.
}


\maketitle

\textbf{Cavity-polaritons in semiconductor microstructures have
emerged as a promising system for exploring nonequilibrium dynamics
of many-body systems~\cite{Cuiti-Carusotto-RMP}. Key advances in
this field, including the observation of polariton
condensation~\cite{Deveaud2006}, superfluidity \cite{Amo09},
realization of topological photonic bands \cite{St-jean17}, and
dissipative phase transitions \cite{Baumberg17, Rodriguez17,
Fink18}, generically allow for a description based on a mean-field
Gross-Pitaevskii formalism. While observation of polariton intensity squeezing~\cite{Karr04, Boulier14} and decoherence of a polarization entangled photon pair by a polariton condensate~\cite{Cuevas18} provide counter-examples, quantum effects in these experiments show up at high polariton occupancy. Going beyond into the regime of strongly
correlated polaritons requires the observation of a photon blockade
effect \cite{Imamoglu97,Ciuti-Carusotto2006} where interactions are
strong enough to suppress double occupancy of a photonic lattice
site. Here, we report the observation of quantum correlations
between polaritons in a fiber cavity which spatially confines
polaritons into an area of 3~$\mu$m$^2$. Photon correlation
measurements show that careful tuning of the coupled system allows
for a modest photon blockade effect as evidenced by a reduction of
simultaneous two-polariton generation probability by 5~\%.
Concurrently, our experiments provide an unequivocal measurement of
the polariton interaction strength, thereby resolving the
controversy stemming from recent experimental reports
\cite{Snoke17}. Our findings constitute a first essential step
towards the realization of strongly interacting photonic systems. }

Photon blockade can be described as strong conditional modification
of  optical properties of a medium by a single photonic excitation.
The early demonstrations of photon blockade were based on the
realization of the Jaynes-Cummings model consisting of a single
anharmonic quantum emitter strongly coupled to a cavity mode. The
requisite optical nonlinearity in this case originates from the
impossibility of double excitation of the anharmonic emitter and
consequently of the cavity-emitter system under resonant excitation.
In the opposite limit of a collective excitation of a large number
of two-level emitters coupled to a cavity mode, photon blockade can
only be observed if there is strong interaction between optically
excited collective states. This for example is manifestly the case
if the incident field creates a collective excitation to a Rydberg
state with large van der Waals interactions~\cite{Cubel05,
Saffman09}. In condensed matter, an analogous situation can be
obtained using quantum well (QW) excitons nonperturbatively coupled
to a cavity mode, leading to the formation of hybrid light-matter
eigenstates, termed polaritons. In this system, exciton-exciton
interactions ensure that the polariton spectrum is anharmonic,
provided that polaritons are confined to maximize their in-plane
spatial overlap. If the interaction-induced energy shift is larger
than the polariton linewidth, the system exhibits polariton
blockade~\cite{Ciuti-Carusotto2006}. A convenient verification of
the blockade phenomenon is provided by photon antibunching -- a
vanishing photon correlation function $g^{(2)}(\tau)$ for delays
$\tau$ smaller than the polariton lifetime. Even in the limit where
the nonlinearity is smaller than the linewidth, finite photon
antibunching can be observed provided that the laser is detuned from
the polariton center frequency. Such an observation confirms a
first-order modification of the cavity transmission conditioned on
the excitation of a single polariton~\cite{Ferretti12}.

\begin{figure*}[h!] 
\centering
\includegraphics[width=0.5\textwidth]{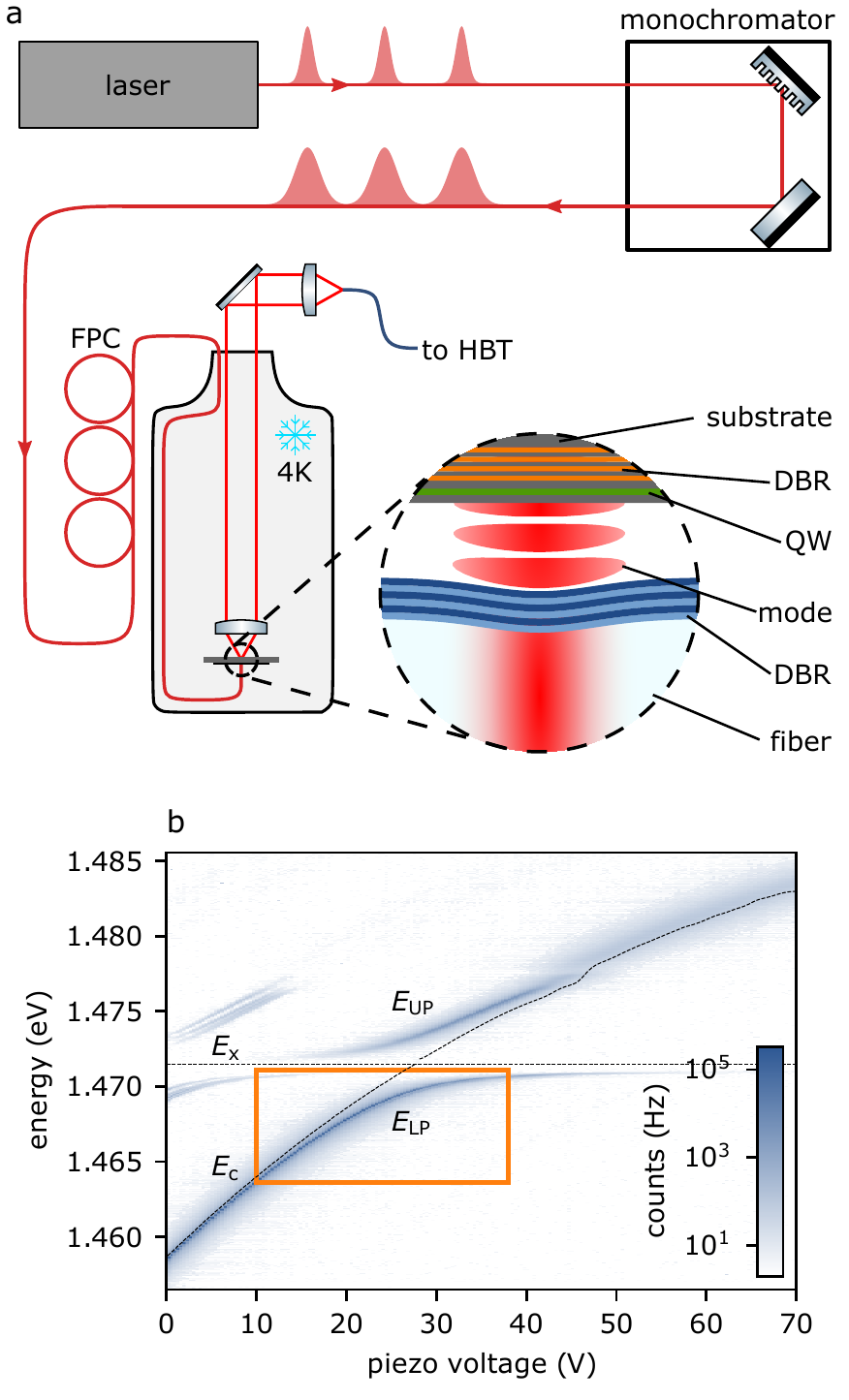}
\caption{{\bf Experimental set-up for cavity-polaritons.} (a) A
fiber cavity is driven by either a cw laser or ps laser pulses whose
pulsewidth is prolonged by a monochromator. The polarization set by fiber polarization controllers (FPC) is chosen to match the lowest linearly polarized cavity mode. The
transmitted light is collected in a fiber and guided to a
Hanbury Brown and Twiss (HBT) setup where photon statistics are measured. Zoom-in: Structure of the semi-integrated cavity. (b)
White light transmission spectra as a function of the piezo voltage
controlling the cavity length, revealing a Rabi splitting of
3.5~meV. The orange box indicates the region of interest for the
present work.} \label{fig01}
\end{figure*}

To demonstrate quantum correlations of confined polaritons, we
measure $g^{(2)}(\tau)$ using photons transmitted from a resonantly
driven zero-dimensional cavity that is strongly coupled to the
excitonic transition of a QW~\cite{Besga15, Fink18}.
Figure~\ref{fig01}a depicts the experimental set-up based on a
semi-integrated fiber cavity whose length is tunable \textit{in
situ} using a piezo-based nanopositioner. The concave surface of the
fiber leads to lateral confinement with a mode waist of $w_0 =
1.0$~$\mu$m. A single InGaAs QW is placed at an antinode of the
fundamental TEM$_{00}$ mode (see Methods). When scanning the cavity
mode energy across the exciton resonance, we observe a large Rabi
splitting of $E_\mathrm{R} = 3.5$~meV between the two polariton
eigenmodes, signaling the strong coupling regime
(Fig.~\ref{fig01}b). Focusing exclusively on the lower polariton
(LP) mode and neglecting the role of the far detuned upper polariton
branch, the Hamiltonian describing the system in the rotating basis
under resonant drive reads:
\begin{equation}
\label{label1} H = (\hbar \omega_\mathrm{L} - E_\mathrm{LP}) \hat{p}^\dagger \hat{p} + \dfrac{U_\mathrm{pp}}{2} \hat{p}^\dagger \hat{p}^\dagger \hat{p} \hat{p} + \hbar F^* \hat{p}^\dagger + \hbar F \hat{p}
\end{equation}
where $\hat{p}$ denotes the lower polariton annihilation operator,
$\omega_\mathrm{L}$ the excitation laser frequency, $E_\mathrm{LP}$ the LP energy, $F$ the
drive strength, and $U_\mathrm{pp}$ the polariton-polariton interaction
strength. The polariton annihilation operator $\hat{p}$ can be decomposed
into its exciton and photon components as $\hat{p} = c_\mathrm{p} \hat{a} + c_\mathrm{x} \hat{x}$, where $\hat{a}$ ($\hat{x}$) denotes  the annihilation operator associated with the cavity mode (QW exciton). The corresponding Hopfield coefficients
$c_\mathrm{p}$ and $c_\mathrm{x}$, satisfying $|c_\mathrm{p}|^2 + |c_\mathrm{x}|^2 = 1$, depend on the cavity-exciton detuning and therefore can be adjusted by changing the cavity length.

\begin{figure*}[h!] 
\centering
\includegraphics[width=\textwidth]{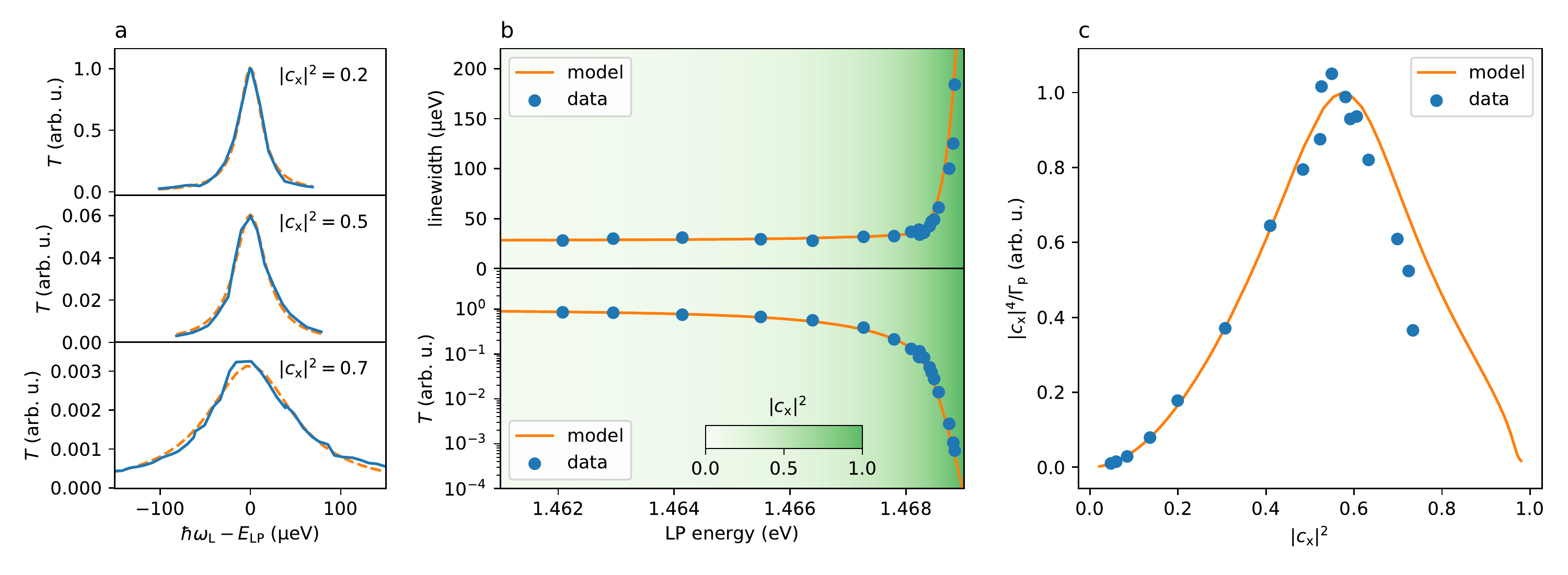}
\caption{{\bf Optimization of polariton parameters}. (a)
Transmission spectra measured at three different cavity-exciton
detunings (blue line) depicted together with the Lorentzian fit to
the data (orange dashed line). (b) Upper panel: polariton linewidth
$\Gamma_\mathrm{p}$ extracted from the transmission measurements, as
a function of the LP energy (blue dots) and linewidth calculated
from the simulation (orange curve). Lower panel: Maximum measured
polariton transmission as a function of the LP energy (blue dots)
and simulated maximum cavity transmission (orange curve). The
background color plot indicates the Hopfield coefficient
corresponding to the lower polariton energy. (c) The ratio
$|c_\mathrm{x}|^4/\Gamma_\mathrm{p}$ which quantifies the maximum
attainable quantum correlations as a function of the exciton content $|c_\mathrm{x}|^2$ of
the lower polariton (blue dots). The calculated ratio
$|c_\mathrm{x}|^4/\Gamma_\mathrm{p}$ (orange curve).} \label{fig02}
\end{figure*}

In the limit of a small nonlinearity, the maximum degree of
antibunching $A \triangleq \max \left( 1 - g^{2}(0) \right)$ is
approximately given by the ratio between the polariton-polariton
interaction energy and the linewidth $\Gamma_\mathrm{p}$: $A \approx U_\mathrm{pp}/\Gamma_\mathrm{p} =
U_\mathrm{xx}|c_\mathrm{x}|^4/\Gamma_\mathrm{p}$ where $U_\mathrm{xx}$ is the exciton-exciton
interaction energy (see Supplementary Information). Thus, the maximum
antibunching is the result of an interplay between polariton
linewidth and exciton content. In the following we identify the
optimal system parameters allowing the observation of blockade by
characterizing the dependence of the lower polariton linewidth and
maximum transmission on the cavity mode energy. We perform laser
scans of the transmission while varying the cavity length (Fig.~\ref{fig02}a),
from which we extract the linewidth and maximum transmission.
Figure~\ref{fig02}b plots the cavity linewidth and transmission as a function
of the LP energy: as we increase $E_\mathrm{LP}$ by reducing the cavity
length, the linewidth experiences a moderate increase up to
$E_\mathrm{LP}\approx 1.468$~eV. For higher $E_\mathrm{LP}$, the linewidth increases
sharply while the corresponding maximum transmission drops.

In order to identify the contributions to the polariton linewidth,
we model the optical response of our cavity with an input-output
formalism~\cite{Diniz11} where we use experimentally measured
parameters: the Rabi splitting $E_\mathrm{R} = 3.5$~meV, the
disorder dominated (bare) exciton linewidth $ h \Delta
\nu_\mathrm{x} = 0.5$~meV measured outside the cavity, and the bare
cavity linewidth $\kappa = 28$~$\mu$eV. The only free parameter in
our model is a Markovian exciton decay rate $\gamma_\mathrm{x}$. We
obtain the best agreement with experimental data for
$\gamma_\mathrm{x} = 40$~$\mu$eV (see Supplementary Information).
Although the physical process associated with $\gamma_\mathrm{x}$ is
unclear, we tentatively attribute this decay channel to disorder
mediated coupling of the confined LP mode to guided modes of the
planar structure. A better understanding of this decay channel would
be beneficial since its elimination could lead to a sizable increase
in the degree of antibunching. On the other hand, the coupling to
localized exciton states has only little effect on the polariton
linewidth as long as we maintain negligible overlap between the LP
mode and the inhomogeneously broadened bare exciton
spectrum~\cite{Houdre}.

This characterization allows to identify the optimal cavity-exciton detuning $\delta_\mathrm{c-x}$ for
maximizing quantum correlations between polaritons, or equivalently,
$A \propto |c_\mathrm{x}|^4/\Gamma_\mathrm{p}$. Figure~\ref{fig02}c shows this ratio calculated for both our data and our model, as a function of the exciton content. We deduce from this plot that
the exciton content at which we expect the highest degree of
antibunching is about $|c_\mathrm{x}|^2 \sim 0.6$, yielding a polariton
lifetime of $\tau_\mathrm{LP} = \hbar /\Gamma_\mathrm{p} \approx 11$~ps. In the following we use the corresponding optimal value of $\delta_\mathrm{c-x}$.

Since $\tau_\mathrm{LP}$ is shorter than the time resolution of
single-photon detectors, we employ a pulsed excitation where the
pulse width $\tau_\mathrm{L}$ of the laser pulses is a factor of
$\approx 3$ longer than $\tau_\mathrm{LP}$~\cite{Munoz18}. The only
requirement on detector bandwidth in this case is that it exceeds
the laser repetition rate of $80$~MHz. This approach allows us to
implement a red-detuned excitation where all laser frequency
components satisfy $\hbar \omega_\mathrm{L} < E_\mathrm{LP}$,
thereby ensuring that $A$ is maximal for repulsive interactions
($U_\mathrm{pp} > 0$). On the other hand, we cannot rule out
multiple excitations during the timescale given by
$\tau_\mathrm{L}$, which reduces the magnitude of the measured $A$.

\begin{figure*}[h!] 
\centering
\includegraphics[width=\textwidth]{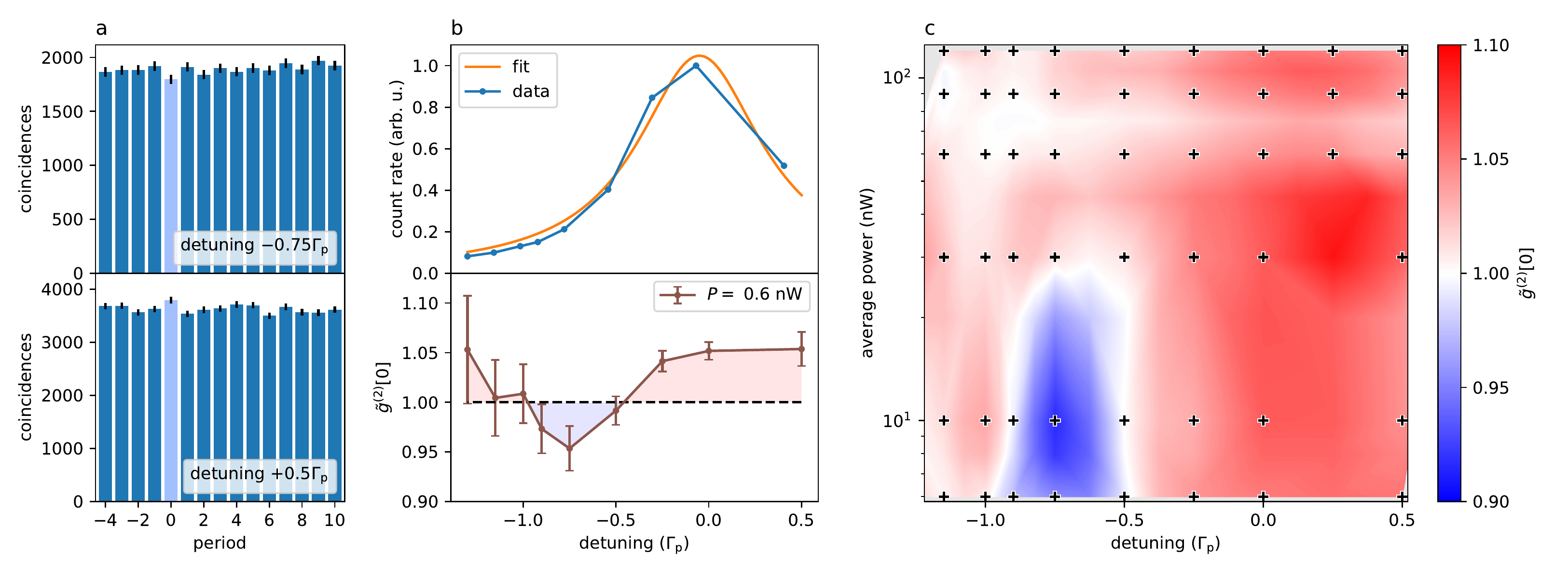}
\caption{{\bf Quantum correlations between polaritons.} (a) Upper
panel: Coincidence number in pulsed regime as a function of the
period difference between the two detected photons, at a detuning of
$\Delta = - 0.75 \Gamma_\mathrm{p}$. Lower panel: Same as the upper panel, but at a
detuning of $\Delta = + 0.5 \Gamma_\mathrm{p}$. (b) Upper panel: Photon rate as a
function of the laser detuning (blue curve). Lorentzian fit to the
data (orange curve). Lower panel: Integrated $\tilde{g}^{2}[0]$ as a
function of the laser detuning. The dashed black line indicates the value for uncorrelated photons. (c) Integrated $\tilde{g}^{2}[0]$ as a function of the laser detuning and power. The black crosses indicate the detuning and power at which the measurements are carried out. The blue areas are below the classical
limit given by $\tilde{g}^{2}[0] \ge 1$. The $\tilde{g}^{2}[0]$
dependence on detuning shown in (b) corresponds to the lowest
depicted power.} \label{fig03}
\end{figure*}

Photons emitted from the cavity are collected in a single mode fiber
and transferred to a Hanbury Brown and Twiss setup where the
coincidences are recorded. The upper (lower) panel of Fig.~\ref{fig03}a shows
the histogram obtained at $\Delta = \hbar \omega_\mathrm{L0} - E_\mathrm{LP} =
-0.75\Gamma_\mathrm{p}$ ($+0.5 \Gamma_\mathrm{p}$), where $\omega_\mathrm{L0}$ denotes the center
frequency of the laser pulse. In the red-detuned case ($\Delta <
0$), we observe that the number of coincidences corresponding to a
period difference of 0 is lower than the average value of the other
periods, while the opposite is true for the blue-detuned case. We
deduce from each histogram the corresponding integrated second order
correlation $\tilde{g}^{2}[0]$ (see Methods).

Figure~\ref{fig03}b  shows $\tilde{g}^{2}[0]$ as a function of the
laser detuning, obtained by repeating such measurements while
scanning the laser energy. We observe a minimum of $\tilde{g}^{2}[0]
= 0.95 \pm 0.02$ for $\Delta = -0.75\Gamma_\mathrm{p}$. The value
for $\Delta = 0$ lies above unity; we tentatively attribute this
bunching signal to feeding of the LP resonance by a thermal
localized exciton bath. Repeating these measurement  for different
incident laser powers allows us to obtain the contour plot in
Fig.~\ref{fig03}c where we show the value of $\tilde{g}^{2}[0]$ as a
function of the laser detuning and power. This plot is constructed
from 63 data points measured with power- and detuning-dependent
integration times, varying from 0.5~h to 8.5~h per point, for a
total measurement time of 82~h. It is evident that photon
antibunching is present for small red detunings up to a power of
$\sim 20$~nW and vanishes at higher excitation power.

The determination of $U_\mathrm{xx}$ has been previously carried out
using the polariton density dependent blue shift of the LP
resonance. Such measurements, however, are inherently inaccurate due
to the difficulty in estimating the actual exciton population which
has contributions not only from polaritons but also from long-lived
localized excitons. The latter are generated even under resonant
excitation, as evidenced by the bunching signal we observe for
$\Delta = 0$. On the other hand, the minimum value of
$\tilde{g}^{(2)}$ directly gives a lower bound on $U_\mathrm{xx}$
without requiring an estimation of the exciton population (see
Supplementary Material). From our measurements we estimate
$U_\mathrm{pp} = 13 \pm 6$~$\mu$eV, yielding $U_\mathrm{xx} =
40$~$\mu$eV$\mu$m$^2$, in agreement with what has been estimated by
several groups~\cite{Ferrier11,Rodriguez16,Rodriguez17,Munoz18}.

Our results demonstrate that the cavity-exciton response to a drive
field can be sizably modified by the presence of a single
polariton. Together with Ref.~\cite{Munoz18}, our observations demonstrate
quantum behavior of polaritons that cannot be captured by mean-field
methods. While the degree of quantum correlations we report are
modest, further enhancement of the degree of antibunching could be
reached by decreasing the polariton linewidth by employing higher
quality QW samples with reduced disorder. Alternatively, $U_\mathrm{xx}$
could be enhanced by using dipolar polaritons in coupled
QWs~\cite{Togan18} embedded in microcavities, or polaron-polaritons
in the fractional quantum Hall regime~\cite{Ravets18}.

\section*{METHODS}

\textbf{Experimental set-up.} The experiments are based on an open
cavity structure where the top mirror is a concave distributed Bragg
reflector (DBR) mirror deposited on a fiber. A dimple with a radius
of curvature of 13.9~$\mu$m is created on the center of the fiber
facet using CO$_2$ laser ablation. The bottom (output) mirror is
integrated in the sample and consists of an AlAs/GaAs DBR grown
by molecular beam epitaxy. The InGaAs QW is grown on top of this
latter DBR during the same growth process. The light transmitted
through the cavity is collimated from the back of the sample and
collected into a fiber that guides it to a Hanbury Brown and Twiss
set-up.

\textbf{Photon correlation measurements.} We drive the cavity  using
a Ti:Sapphire picosecond laser with 80~MHz repetition rate. We prolong
the pulses using a monochromator of bandwidth 20~$\mu$eV. The center
wavelength of the laser pulses is set by controlling the grating
angle. The transmitted photons are detected by two fiber-based
commercial avalanche photodiodes of quantum efficiency 45~\% and timing jitter 300~ps, and the detection events are
recorded using a time-tagged single photon counter. The photon
records are binned according to the number of laser periods $\Delta
T$ separating their detection. The integrated $\tilde{g}^{(2)}[0]$ is
defined as the ratio between the coincidence number with $\Delta T =
0$ to the average of the coincidence number having $\Delta T\neq0$
calculated from 20~consecutive periods around the zero delay.

{}

\vspace{1 cm}

\textbf{Supplementary Information} is linked to the online version of the paper at www.nature.com/nature.\\

\textbf{Acknowledgments} This work was supported by the Swiss National Science Foundation (SNSF) through a DACH project 200021E-158569-1, SNSF National Centre of Competence in Research - Quantum Science and Technology (NCCR QSIT) and an ERC Advanced investigator grant (POLTDES). The W\"urtzburg Group acknowledges support by the state of Bavaria, and the DFG within the project SCHN1376-3.1\\

\textbf{Author Contributions} A.D., T.F. and A.I. supervised the
project. T.F. designed the experiment. A.D. and T.F. carried out the
measurements. A.S., C.S., and
S.H. grew the sample. A.D. and A.I. wrote the manuscript.\\

\textbf{Author Information} Reprints and permissions information is available at www.nature.com/reprints. The authors declare that they have no competing financial interests. Correspondence and requests for materials should be addressed to imamoglu@phys.ethz.ch.\\

\newpage

\newpage

\newpage

\end{document}